\documentclass{aastex62}
\usepackage{amssymb}
\usepackage{lineno}
\usepackage{amsmath}
\usepackage{gensymb}
\usepackage{graphicx}
\usepackage{xcolor}

\newcommand{\beq}{\begin{equation}}
\newcommand{\eeq}{\end{equation}}

\newcommand{\Ms}{\textrm{M}_*}
\newcommand{\Msun}{\textrm{M}_\odot}
\newcommand{\kmps}{km~s$^{-1}$}
\newcommand{\MHI}{\rm{M_{H{\textsc i}}}}

\newcommand{\MB}{{\rm M_B}}

\newcommand{\htwo}{{\rm H_2}}

\newcommand{\hi}{H{\sc i}}
\newcommand{\sigc}{\sigma_{\rm cont}}
\newcommand{\sigl}{\sigma_{\rm HI}}
\newcommand{\hii}{H{\sc i} 21\,cm}
\defcitealias{NatPaper}{vD18}
\graphicspath{{./}{figures/}}

\shorttitle{H{\sc i} 21\,cm Emission at $z \approx 1.3$}
\shortauthors{Chowdhury et al.}

\begin{document}
	
	\title{Giant Metrewave Radio Telescope Detection of H{\sc i} 21\,cm Emission from Star-forming Galaxies at $z \approx 1.3$}

	\correspondingauthor{Aditya Chowdhury}
	\email{chowdhury@ncra.tifr.res.in}
	
	\author{Aditya Chowdhury}
	\affil{National Centre for Radio Astrophysics, Tata Institute of Fundamental Research, Pune, India.}
	
	\author{Nissim Kanekar}
	\affil{National Centre for Radio Astrophysics, Tata Institute of Fundamental Research, Pune, India.}
	
	\author{Barnali Das}
	\affil{National Centre for Radio Astrophysics, Tata Institute of Fundamental Research, Pune, India.}
	
	\author{K.S. Dwarakanath}
	\affil{Department of Astronomy and Astrophysics, Raman Research Institute, Bangalore, India. }
	
	\author{Shiv Sethi}
	\affil{Department of Astronomy and Astrophysics, Raman Research Institute, Bangalore, India. }

	
	
	\begin{abstract}
		We report a $\approx 400$-hour Giant Metrewave Radio Telescope (GMRT) search for H{\sc i} 21\,cm emission from star-forming galaxies at $z = 1.18-1.39$ in seven fields of the DEEP2 Galaxy Survey. Including data from an earlier 60-hour GMRT observing run, we co-added the H{\sc i} 21\,cm emission signals from 2,841 blue star-forming galaxies that lie within the full-width at half-maximum of the GMRT primary beam. This yielded a $5.0\sigma$ detection of the average H{\sc i} 21\,cm signal from the 2,841 galaxies at an average redshift $\langle z \rangle \approx 1.3$, only the second detection of H{\sc i} 21\,cm emission at $z\ge1$. We obtain an average H{\sc i} mass of $\langle {\rm M_{HI}} \rangle=(3.09 \pm 0.61) \times 10^{10}\  {\rm M}_\odot$ and an H{\sc i}-to-stellar mass ratio of $2.6\pm0.5$, both significantly higher than values in galaxies with similar stellar masses in the local Universe. We also stacked the 1.4~GHz continuum emission of the galaxies to obtain a median star-formation rate (SFR) of $14.5\pm1.1\ \Msun \textrm{yr}^{-1}$.
		This implies an average H{\sc i} depletion timescale of $\approx 2$~Gyr for blue star-forming galaxies at $z\approx 1.3$, a factor of $\approx 3.5$ lower than that of similar local galaxies. Our results suggest that the \hi\ content of galaxies towards the end of the epoch of peak cosmic SFR density is insufficient to sustain their high SFR for more than $\approx 2$~Gyr. Insufficient gas accretion to replenish the H{\sc i} could then explain the observed decline in the cosmic SFR density at $z< 1$.
	\end{abstract}
	
	
	\section{Introduction}
	
	Understanding galaxy evolution requires us to understand the evolution of, and the interplay between, the two main 
	baryonic components of galaxies, the stars and the interstellar medium (ISM). For most galaxies in the local Universe, 
	the dominant component of the ISM, by mass, is neutral atomic hydrogen (\hi), the primary fuel for star-formation.
	The \hi\ mass of different types of galaxies, and the relations between the \hi\ mass and the molecular gas mass, the stellar mass, and the 
	star-formation rate (SFR), are thus critical inputs to studies of galaxy evolution. In the local Universe, the 
	\hi\ mass of galaxies has long been measured via emission studies in the \hii\ emission line. Unfortunately, the low Einstein 
	A-coefficient of this transition has meant that it has not been possible to detect \hii\ emission from individual 
	galaxies beyond even fairly low redshifts, $z \gtrsim 0.4$, with the highest-redshift detection till date at 
	$z = 0.376$ \citep[e.g.][]{fernandez16}.
	
	Over the last two decades, the contrast between the dramatic improvement in measurements of the stellar properties of 
	high-$z$ galaxies \citep[e.g.][]{Madau14} and the lack of information on their \hi\ properties has become more and 
	more stark. For example, it is now well known that the SFR density of the Universe increases steadily from 
	$z \approx 8$ to $z \approx 3-4$, remains flat from $z \approx 3$ to $z \approx 1$, and then declines precipitously, by 
	an order of magnitude, from $z \approx 1$ to the present epoch \citep[e.g.][]{LeFloch05,Hopkins06,Bouwens10}. The causes
	of this evolution, especially the steep decline in the SFR density after $z \approx 1$, remain unclear today. Measurements 
	of the fuel for star-formation, the \hi\ content of galaxies, especially during the peak epoch of star-formation in the 
	Universe, $z \approx 1-3$, are critical to understanding the evolution of the SFR density.
	
	While measurements of the \hi\ masses of individual galaxies at $z \approx 1$ would require prohibitively large 
	integrations with today's radio telescopes, progress can be made by ``stacking'' the \hii\ emission signals 
	of a large number of galaxies with known spectroscopic redshifts, that lie within the primary beam of a radio 
	interferometer \citep{Chengalur01,Zwaan00}. Such \hii\ stacking yields a measurement of the {\it average} 
	\hi\ content of the stacked galaxies, and thus allows one to obtain statistical information about 
	the \hi\ properties of different galaxy populations at the redshift of interest. Until recently, even such \hii\ 
	stacking studies were mostly limited by sensitivity and frequency coverage to relatively low redshifts, $z \lesssim 0.4$ 
	\citep[e.g.][]{Lah07,Rhee16,Bera19}, with a single search at $z > 1$ yielding a non-detection of the 
	stacked \hii\ emission signal and an upper limit to the average \hi\ mass of a sample of star-forming galaxies 
	at $z \approx 1.3$ \citep{Kanekar16}.
	
	Recently, \citet{Chowdhury20} used the new $550-850$~MHz receivers and the new wideband correlator of 
	the upgraded Giant Metrewave Radio Telescope (GMRT) to obtain the first detection of the stacked \hii\ emission signal 
	at $z \approx 1$. They stacked the \hii\ emission signals from 7,653 star-forming galaxies at $z = 0.74 - 1.45$ in the 
	DEEP2 Galaxy Survey fields \citep{Newman13}, to measure the average \hi\ mass of star-forming galaxies 
	at $z \approx 1$. In this {\it Letter}, we report an independent $\approx 400$-hour GMRT search for \hii\ emission from a sample 
	of star-forming galaxies at $z\approx1.18-1.39$, also in the DEEP2 Survey fields, carried out using the original GMRT 610~MHz 
	receivers and the legacy narrow-band GMRT correlator.

	\section{Observations and Data Analysis}
	
	\begin{table*}[t!]
		\centering
		\caption{Details of the observations: The first seven rows of the table describe the $\approx 400$-hour GMRT observations presented in this {\it Letter}, while the last four rows are for the data of \citet{Kanekar16}. The columns are (1)~the DEEP2 sub-field targetted by the GMRT pointing, (2-3)~the J2000 co-ordinates of the pointing centre, (4)~the on-source time for each frequency setting, where setting~\textbf{A} covers $591.0-624.3$~MHz, setting~\textbf{B} covers $616.0-649.3$~MHz, setting~\textbf{C} covers $601.0-634.3$~MHz, and setting~\textbf{D} covers $626.0-659.3$~MHz (note that a single frequency setting was used for each pointing by \citet{Kanekar16}), (5)~the RMS noise on the continuum image, 
			in $\mu$Jy/beam, (6)~the synthesized beam of the continuum image, (7)~the final number of galaxies used for the stacking analysis from each frequency setting and each GMRT pointing, (8)~the median RMS noise per 34~\kmps\ channel on the \hii\ spectra of the final sample of galaxies in each frequency setting and GMRT pointing, after correcting for their location in the GMRT primary beam.
		}
		\label{table:obs}
		\begin{tabular}{|c|c|c|c|c|c|c|c|c|c|c|}
			\hline
			Field      & Right Ascension & Declination  &    \multicolumn{2}{c|}{On-source} & $\sigc$ & Beam & \multicolumn{2}{c|}{ Number of} & \multicolumn{2}{c|}{$\sigl$}\\
			&   (J2000)  & (J2000)   &  \multicolumn{2}{c|}{Time (hr)} & $\mu$Jy/beam & (continuum) & \multicolumn{2}{c|}{Galaxies } & \multicolumn{2}{c|}{$\mu$Jy/beam}            \\
			& & & \textbf{A} & \textbf{B} & & & \textbf{A} & \textbf{B} & \textbf{A} & \textbf{B} \\
			\hline
			21 & 16h47m54.00s & $34\degree56'00.0"$ & 24.6 & 23.9 & 12 & $4.5''\times 3.8''$ & 103 & 122 & 239 & 184 \\
			22 & 16h50m54.00s & $34\degree56'00.0"$ & 16.8 & 16.4 & 17 & $4.4''\times 3.8''$ & 147 & 142 & 296 & 342 \\
			31 & 23h27m00.00s & $00\degree07'00.0"$ & 18.5  & 13.6 & 12 & $5.4''\times 4.0''$ & 148 & 154 & 250 & 280 \\
			32 & 23h29m30.00s & $00\degree10'00.0"$ &  21.9 & 19.5 & 11 & $5.3''\times 4.0''$ & 183 & 128 & 228 & 210\\
			33 & 23h33m00.00s & $00\degree07'00.0"$ & 15.0 & 19.5 & 13 & $5.5''\times 3.9''$ & 159 & 136 & 281 & 175\\
			41 & 02h27m30.00s & $00\degree35'00.0"$ & 15.7 & 17.1 & 15 & $5.3''\times 3.9''$& 188 & 204 & 288 & 316\\
			42 & 02h30m00.00s & $00\degree35'00.0"$ & 19.7 & 20.3 & 14 & $5.5''\times 3.8''$ & 92 & 262 & 213 & 265\\
			\hline \hline
			21    & 16h47m54.00s & $34\degree56'00.0"$ & \multicolumn{2}{c|}{8.5 \textbf{(C)}} & 16$^{a}$ & $6.7''\times 5.2''$ & \multicolumn{2}{c|}{146} & \multicolumn{2}{c|}{323}\\
			22    & 16h50m54.00s & $34\degree56'00.0"$ & \multicolumn{2}{c|}{8.5 \textbf{(C)}} & 18$^{a}$ & $5.2''\times 4.3''$ & \multicolumn{2}{c|}{154} & \multicolumn{2}{c|}{259}\\
			31/32 & 23h28m00.00s & $00\degree09'00.0"$ & \multicolumn{2}{c|}{13.0 \textbf{(D)}}  & 17$^{a}$ & $5.9''\times 4.6''$ & \multicolumn{2}{c|}{185} & \multicolumn{2}{c|}{309}\\
			32/33 & 23h32m00.00s & $00\degree09'00.0"$ & \multicolumn{2}{c|}{13.0 \textbf{(D)}}  & 13$^{a}$ & $6.1''\times 4.4''$ & \multicolumn{2}{c|}{188} & \multicolumn{2}{c|}{289}\\
			\hline
		\end{tabular}
		\vskip 0.1in
		$^a$The continuum images are from \citet{Bera18}.
	\end{table*}
	
	\begin{figure*}
		\centering
		\includegraphics[width=0.23\linewidth]{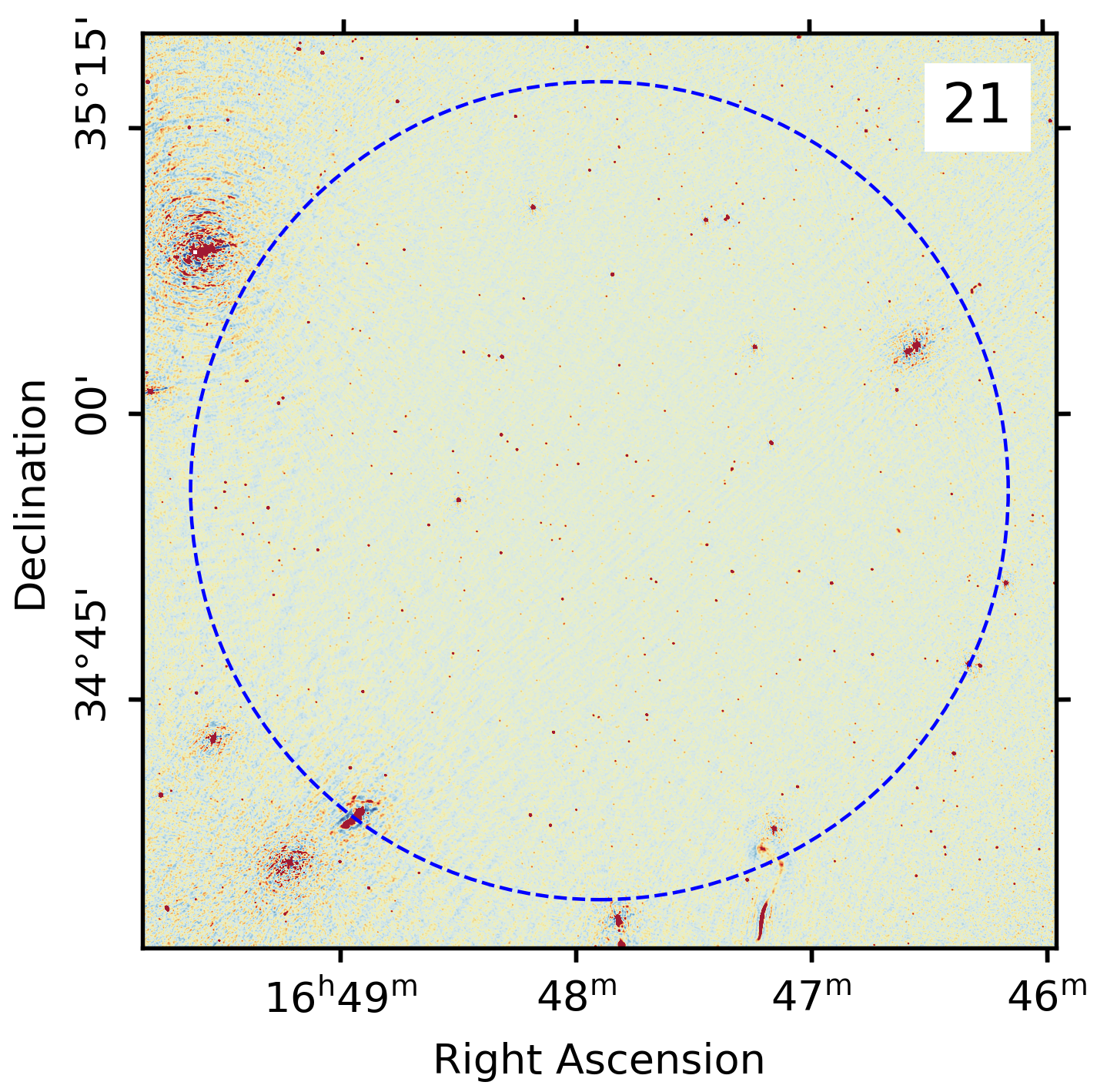}
		\includegraphics[width=0.23\linewidth]{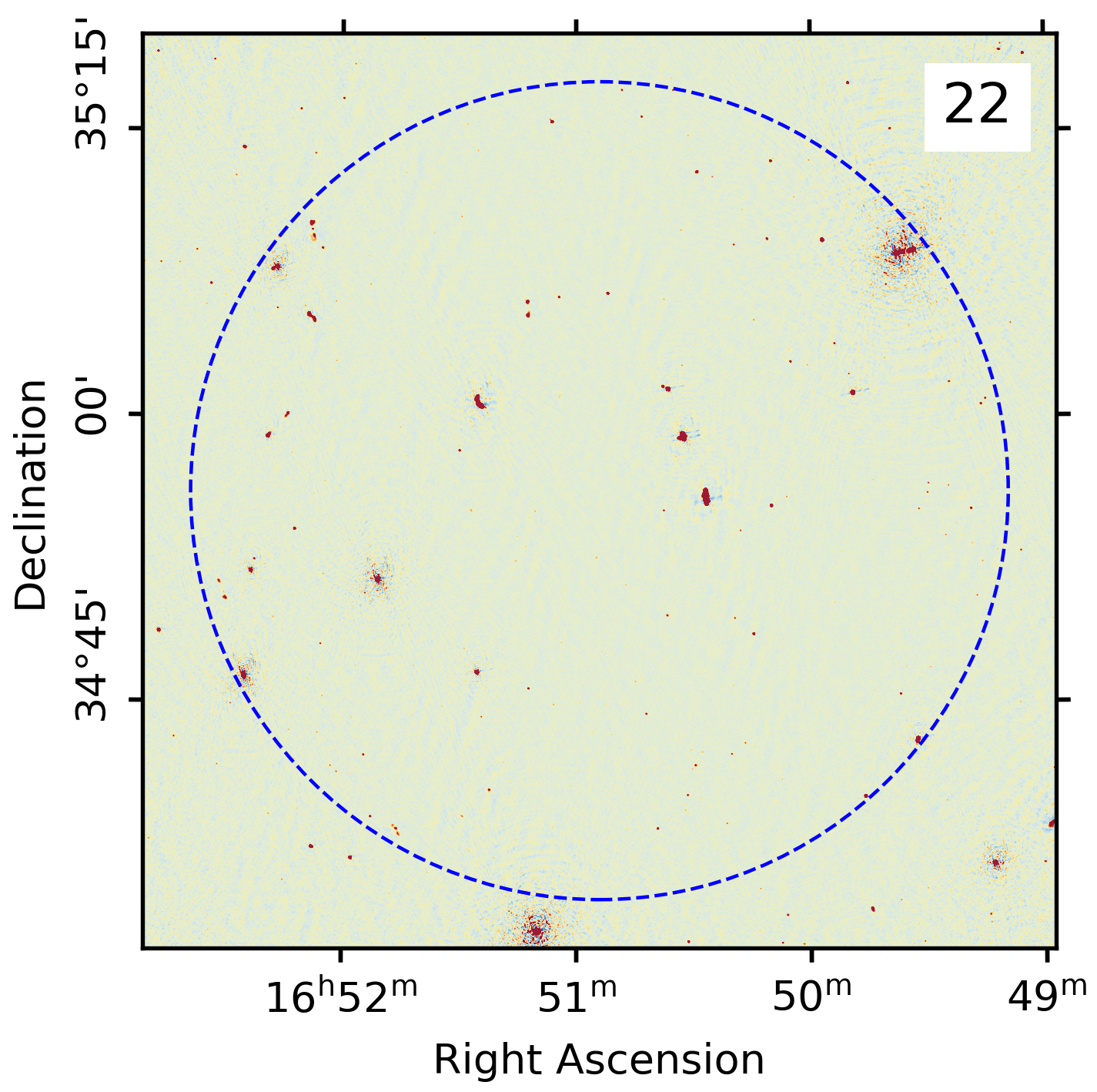}
		\includegraphics[width=0.23\linewidth]{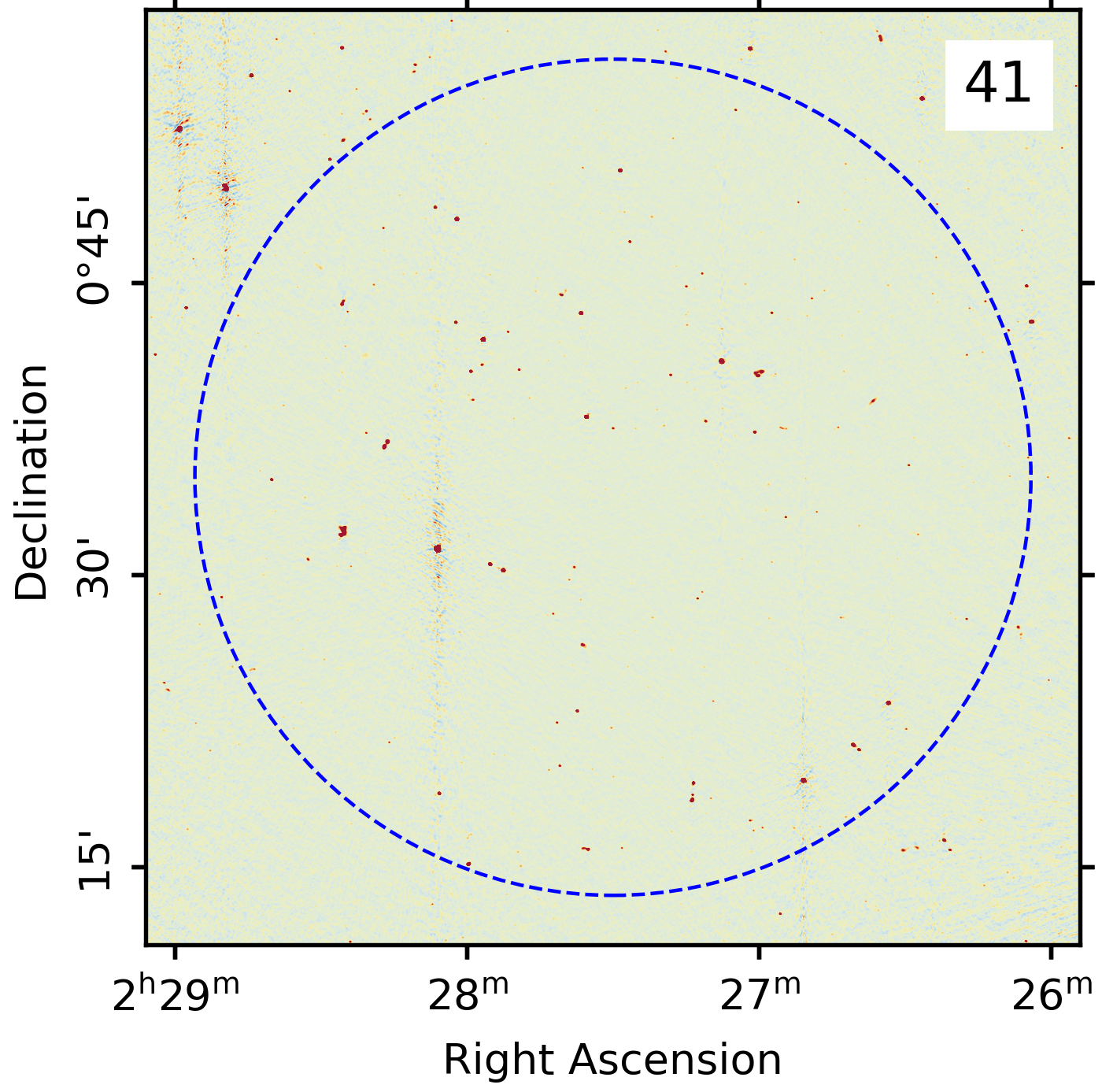}
		\includegraphics[width=0.23\linewidth]{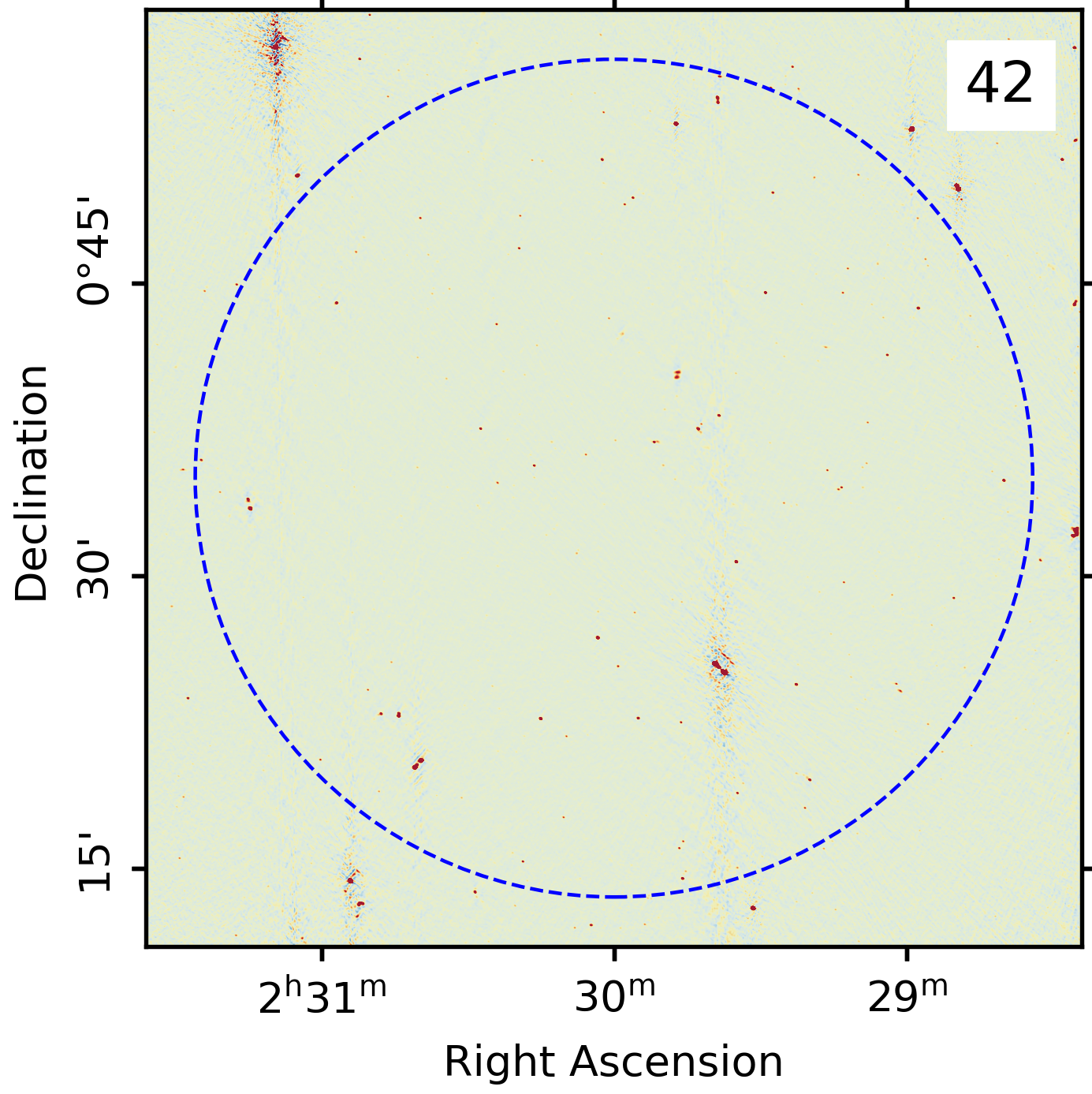}
		
		\includegraphics[width=0.23\linewidth]{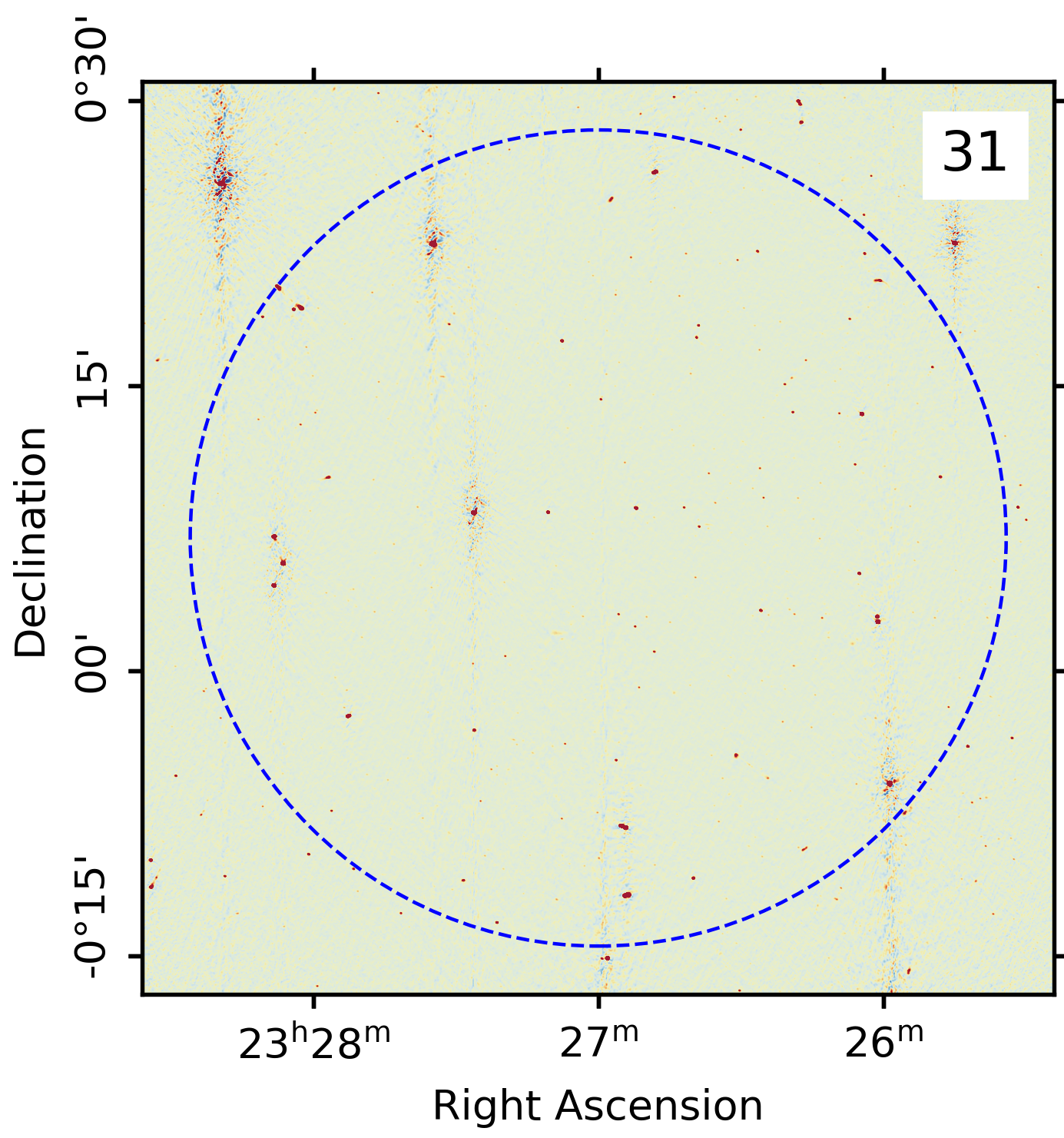}
		\includegraphics[width=0.23\linewidth]{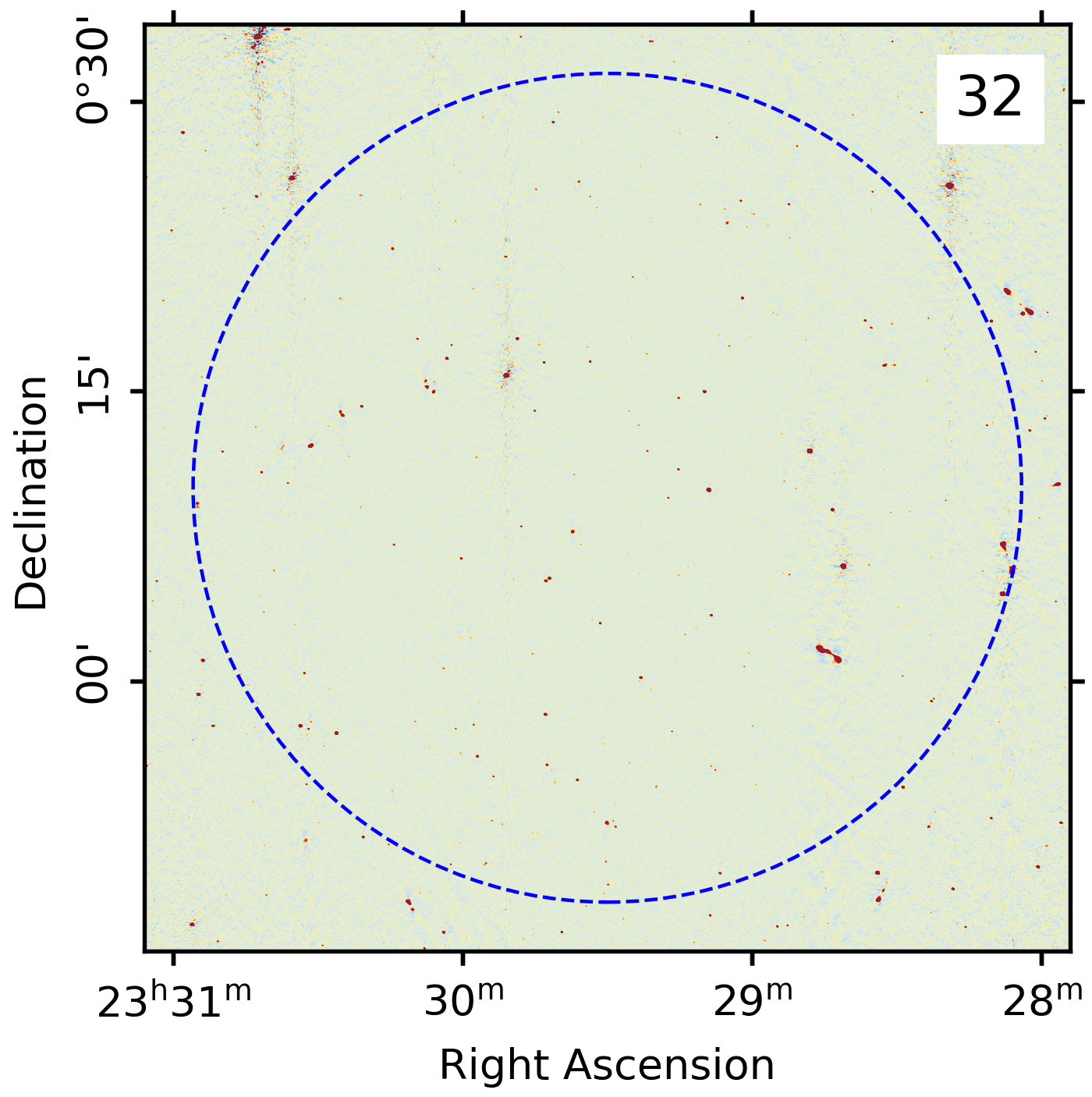}
		\includegraphics[width=0.23\linewidth]{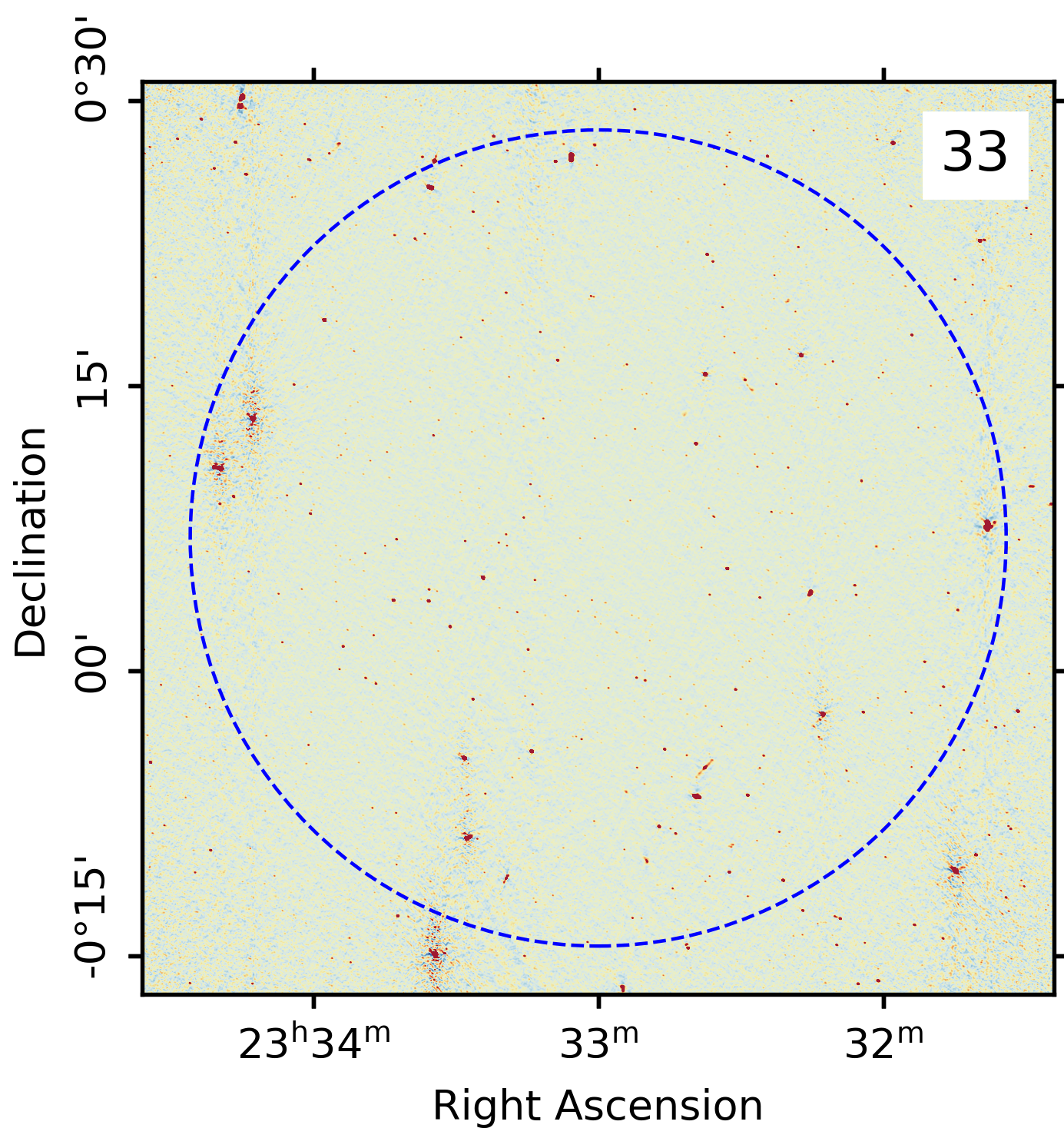}
		\caption{The GMRT 610~MHz continuum images of the seven DEEP2 sub-fields from the $\approx 400$-hours of data presented here. The blue dotted circle in each panel marks the FWHM of the GMRT primary beam at $610$~MHz.}
		\label{fig:cont}
	\end{figure*}

	The \hii\ stacking technique requires a large number of galaxies with accurate spectroscopic redshifts \citep[velocity error, $\Delta V\lesssim 100$~\kmps; e.g.][]{Maddox13} within the primary beam of the interferometer, and with redshifts such that the \hii\ line is redshifted into the frequency band of the telescope. The DEEP2 galaxy redshift survey \citep{Newman13} provides spectroscopic redshifts with $\Delta V\lesssim55$~\kmps\ for galaxies at $z\approx0.7-1.5$, in fields that can be optimally covered by the GMRT primary beam at $\approx 610$~MHz \citep[e.g.][]{Kanekar16,Chowdhury20}. This made the DEEP2 fields the ideal targets for our GMRT \hii\ emission survey at $z\approx1.3$. We used one GMRT pointing to cover each of the seven DEEP2 sub-fields 21, 22, 31, 32, 33, 41, and 42 \citep{Newman13}.
	
	We observed the DEEP2 fields with the original GMRT 610~MHz receivers for a total of $\approx 400$-hrs between 2015 May and 2016 November (project codes:  $28\_097$, $29\_034$, $30\_069$, $31\_118$; P.I: N. Kanekar). We used the GMRT Software Backend (GSB) as the correlator, with a 33.3~MHz observing band, sub-divided into 512~spectral channels. For each pointing, we used two frequency settings, $591.0-624.3$~MHz and $616.0-649.3$~MHz, to cover the \hii\ line from the redshift range $z=1.19-1.39$, with a velocity resolution of  $\approx 30-33$~\kmps. We carried out seven pointings on the DEEP2 fields, with a total time of $\approx 40-60$~hours per pointing, split approximately evenly between the two frequency settings. Observations of one or more of 3C286, 3C48 and 3C147 in each observing run were used to calibrate the flux density scale, while observations of the compact sources 0022+002, 0204+152, or 1609+266 were used to calibrate the complex antenna gains and bandpass shapes. The total on-source time for each pointing, combining the two frequency settings, was $\approx 32-49$ hours. The observational details are summarized in Table~\ref{table:obs}.
	
	All data were analyzed in the Common Astronomy Software Applications Package \citep[{\sc casa} version~5; ][]{McMullin07}, following standard procedures. The {\sc aoflagger} package \citep{Offringa12} was additionally used for excision of data affected by radio frequency interference (RFI). For each pointing, after initial excision of non-working antennas and data affected by RFI, and calibration of the antenna gains and bandpasses, the target visibilities of the two frequency settings were combined, and a standard self-calibration procedure carried out on the combined data set. The antenna-based complex gains and bandpasses at all stages of the analysis were determined using a custom package within the {\sc casa} framework; the complex gain solver in this package is robust to the presence of outliers in the data\footnote{The package is publicly available at \url{https://github.com/chowdhuryaditya/calR} and is archived in Zenodo \citep{calR}.}. We note that, at the time when the data were being analyzed, the {\sc casa} calibration routines did not contain robust solvers, and were hence unstable in the presence of RFI, typical at these low radio frequencies. Imaging was done using the {\sc casa} task {\sc tclean}, with w-projection \citep{Cornwell08}, and MT-MFS (first order expansion; \citealp{Rau11}); we used Briggs weighting with a robust parameter of -0.5 \citep{Briggs95} during self-calibration and a robust parameter of 0.0 for the final continuum imaging. For each target field, a region of size $\approx 1.8\degree \times 1.8\degree$ was imaged, extending far beyond the null of the GMRT primary beam at these frequencies. The final continuum images obtained after self-calibration are shown in Figure~\ref{fig:cont}; the images have an RMS noise of $\approx 11-17 \ \mu$Jy/beam, away from radio continuum sources, with a synthesized beam of $\approx 3.8''-5.5''$ (see Table~\ref{table:obs}).
	
	Next, the continuum emission of each field was subtracted from the self-calibrated visibilities using the tasks {\sc uvsub}, followed by {\sc mstransform}; the latter task was also used to regrid the visibilities to the barycentric velocity frame. Following this, the continuum-subtracted and regridded visibilities of each pointing and frequency setting were used to make the final spectral cubes; this was done using w-projection and Briggs weighting with a robust parameter of +1 \citep{Briggs95}. We experimented with different choices of the robust parameter, and found that a robust parameter of +1 suppresses the spatial extent of deconvolution errors around bright radio-continuum sources while having a negligible effect on the RMS noise of the cube. The final spectral cubes have an angular resolution of $\approx 5.3''-8.0''$, corresponding to spatial resolutions of $\approx 46\ \textrm{kpc}-67\ \textrm{kpc}$ at $z \approx 1.3$.\footnote{Throughout this work, we use a flat $\Lambda$-cold dark matter ($\Lambda$CDM) cosmology, 
		with ($\rm H_0$, $\rm \Omega_{m}$, $\rm \Omega_{\Lambda})=(70$~km~s$^{-1}$~Mpc$^{-1}$, $0.3, 0.7)$.}
	
	Besides the $\approx 400$-hours of new GMRT data, we also made new spectral cubes (again with w-projection and a robust parameter of +1) from the self-calibrated and continuum-subtracted visibilities of the 60-hour data set of \citet{Kanekar16}. These observations also used the GMRT 610-MHz receivers and the GSB as the correlator, with a 33.3~MHz bandwidth and 512~channels. Four GMRT pointings were observed, each with a single frequency setting, covering $601.0-634.3$~MHz or $626.0-659.3$~MHz. For completeness, we present a summary of these observations in Table~\ref{table:obs}.
	
	We corrected for any offset in the astrometry of our images using our recent upgraded GMRT images of these fields \citep[e.g.][]{Chowdhury20}; the astrometry of the latter images has been verified to be consistent with that of the DEEP2 survey \citep{Newman13}. In all cases, the positional offset of the images was comparable to or lower than the full-width-at-half maximum (FWHM) of the synthesized beam of the final continuum image.
	
	\section{Stacking the \hii\ emission signals}
	
	The DEEP2 DR4 catalog contains 3,109 galaxies with accurate redshifts \citep[Redshift Quality Code, Q$\ge3$; ][]{Newman13}, for which the redshifted \hii\ line frequency lies within our frequency coverage, and that lie within the FWHM of the GMRT primary beam at the redshifted \hii\ line frequency. However, the \hii\ lines of some of these 3,109 galaxies were covered multiple times, due to their location in the overlap region of the seven new GMRT pointings, or due to the galaxies being covered in the GMRT pointings of both the $\approx 400$-hour new data set and the observations of \citet{Kanekar16}. Accounting for the multiple observations, we have a total of 3,996 independent \hii\ spectra; for simplicity, we will hereafter refer to each of these spectra as arising from individual galaxies (i.e. will ignore the fact that some of the spectra are from the same object). Of the 3,996 ``galaxies'', 3,142 are covered by the $\approx 400$-hours of new data, while 854 galaxies are from \citet{Kanekar16}.  
	
	We imposed further restrictions on the above sample of 3,996 galaxies to ensure homogeneity of the sample and good data quality. The selection criteria as well as the procedure to stack the \hii\ emission of the final sample are similar to those used by \citet{Chowdhury20}. First, the DEEP2 selection criterion \citep[R~$\leq 24.1$;][]{Newman13} preferentially picks out blue star-forming galaxies at $z \gtrsim 1$ \citep[e.g.][]{Weiner09}. To ensure homogeneity in our sample, we excluded the 162 galaxies which are part of the ``red cloud" in the color-magnitude diagram \citep{Willmer06}. We further excluded 238 radio-loud AGNs from our sample, in order to restrict ourselves to star-forming galaxies; the AGNs were identified based on their detection at $\geq 4\sigma$ significance in our continuum images, with a rest-frame 1.4~GHz luminosity $> 2\times 10^{23}$~W~Hz$^{-1}$ \citep{Condon02}\footnote{We note that our earlier papers \citep[e.g.][]{Bera19,Chowdhury20}, the same quantity was referred to as ``luminosity density''; here we use  ``luminosity'' throughout, to be consistent with the usage in radio astronomy}. After excluding red galaxies and AGNs, our sample contains 3,596 blue star-forming galaxies, at $z = 1.18-1.39$.
	
	
	We next extracted three-dimensional sub-cubes around each of the 3,596 galaxies, and convolved the sub-cubes to a uniform beam of $60$~kpc at the redshift of each galaxy. At this stage, we excised any spectral channels whose intrinsic synthesized beam is $>60$~kpc. We then regridded each convolved sub-cube to the same spatial and spectral grid, with spatial pixels of size $5.2$~kpc, covering  $\pm 260$~kpc around each galaxy, and velocity channels of $34$~\kmps, covering  $\pm1500$~\kmps\ around the galaxy's redshifted \hii\ line frequency. We then fitted a second-order spectral baseline to each spatial pixel of all sub-cubes, and subtracted out this baseline; this removes systematic effects caused by deconvolution errors and bandpass inaccuracies. 
	
	Next, we inspected the \hii\ spectra at the locations of the 3,596 galaxies for any systematic effects that might limit the RMS noise on the final stacked \hii\ spectrum. This was done by testing the spectra for Gaussianity, following the approach of \citet{Chowdhury20}. The p-value thresholds chosen for the Gaussianity tests are slightly different from those in \citet{Chowdhury20}; however, the exact p-value thresholds do not have a significant impact on our results. After excluding 755 galaxies that fail the Gaussianity tests, our sample contains 2,841 blue star-forming galaxies at $z=1.18-1.39$, of which 2,168 galaxies are from the new data presented here, with the remaining 673 galaxies from the data set of \citet{Kanekar16}. The last four columns of Table~\ref{table:obs} list, for each pointing and frequency setting, the number of galaxies and the median RMS noise per $34$~\kmps\ channel on the \hii\ spectra.
	
	For each sub-cube, we converted the \hii\ flux density (${\rm S_{HI}}$) to the \hii\ line luminosity (${\rm L_{HI}}$), using the relation ${\rm L_{HI}}=4\pi \ 
	{\rm S_{HI}} \ {\rm D_L}^2/(1+z)$, where ${\rm D_L}$ is the luminosity distance of the galaxy. We then stacked, pixel by pixel, the three-dimensional \hii\ sub-cubes of each of the 2,841 galaxies. While stacking, we weighted each galaxy by the inverse of the RMS noise on its \hii\ spectrum, in luminosity units. Finally, we fitted a second-order spectral baseline to each pixel of the stacked cube, excluding the central $\pm250$~\kmps, and subtracted out this baseline. 
	
	We estimated the RMS noise on the final stacked \hii\ cube via a Monte Carlo simulation, where we randomly shifted the velocity of the individual galaxies within $\pm1500$~\kmps, and then stacked the velocity-shifted sub-cubes, to obtain a realization of a stacked noise cube. We performed 10,000 realizations of this simulation to determine the RMS noise on the final stacked \hii\ cube. 
	
	\begin{figure*}
		\centering
		\includegraphics[width=\linewidth]{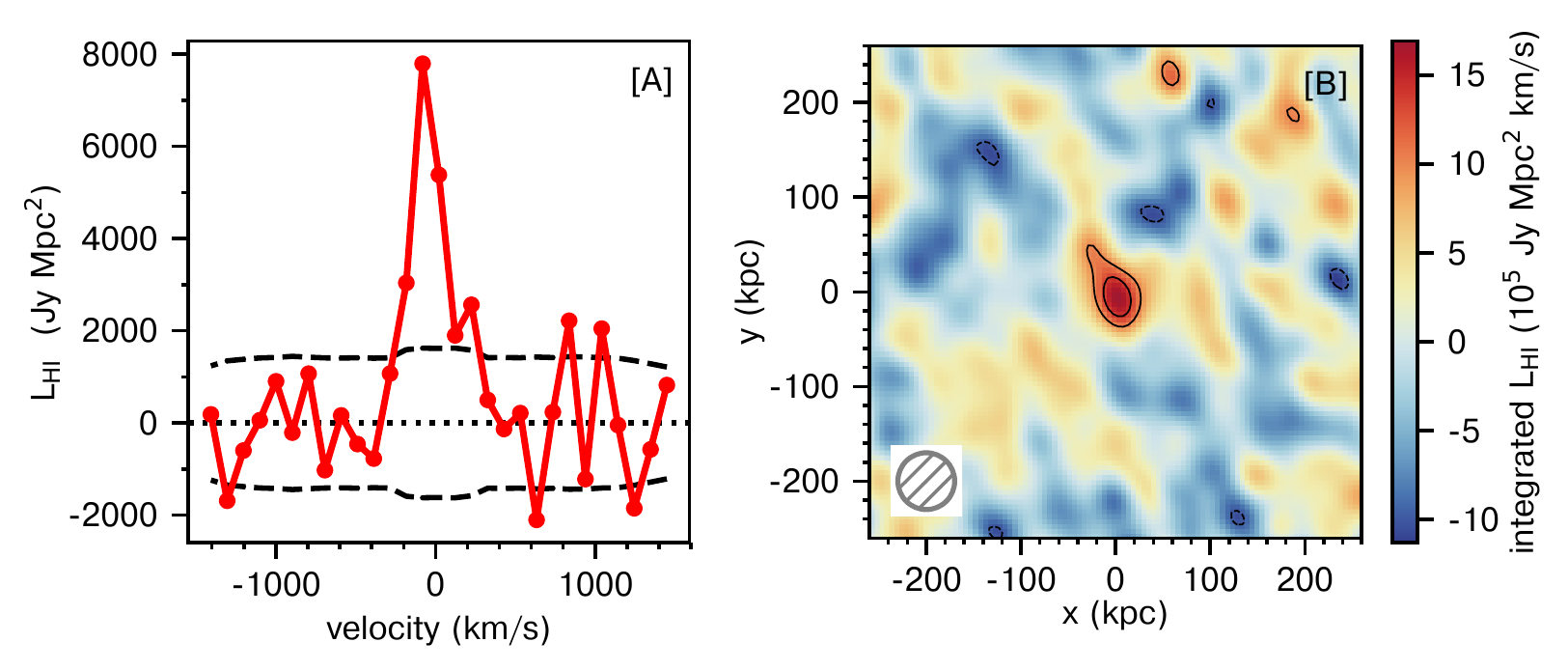}
		\caption{[A] The average \hii\ spectrum, at a velocity resolution of $102$~\kmps, obtained after stacking the individual \hii\ spectra at the location of the 2,841 blue star-forming galaxies at $z=1.18-1.39$. The dashed lines show the $1\sigma$ uncertainty on the spectrum. A clear detection of the average \hii\ signal can be seen in the central velocity channels. [B] The average \hii\ emission image obtained by combining the central channels of the stacked cube. The circle at the bottom left indicates the 60~kpc resolution of the image. The contours are at $(-3, 3, 4.2) \times \sigma$ levels, dashed negative contours. A clear detection of the average \hii\ emission is obtained at the centre of the image.}
		\label{fig:signal}
	\end{figure*}

	The final stacked \hii\ spectrum, obtained by taking a cut through the center of the stacked cube (i.e. at the position of the stacked galaxies), is shown, at a velocity resolution of $102$~\kmps, in Figure~\ref{fig:signal}[A]. The dotted line in the figure shows the $\pm1\sigma$ uncertainty derived from the  Monte Carlo simulation described above. We obtain a clear detection, with $5.0\sigma$ statistical significance, of the average \hii\ emission signal from the 2,841 galaxies at an average redshift of $\langle z \rangle \approx 1.3$.  Figure~\ref{fig:signal}[B] shows the velocity-integrated \hii\ emission image, obtained by adding the central three velocity channels of the final stacked \hii\ cube. The image also shows a clear detection of the \hii\ emission signal in the central region. 
	Integrating over the stacked \hii\ line profile yields a  velocity-integrated \hii\ line luminosity of $(16.5\pm3.3)\times10^{5}$ Jy~Mpc$^2$~\kmps. This corresponds to an average \hi\ mass of $\langle \MHI \rangle=(3.09\pm0.61)\times10^{10}\ \Msun$, for the 2,841 blue star-forming galaxies at  $\langle z \rangle \approx 1.3$.

	We tested whether the \hii\ emission is extended on scales larger than 60~kpc by carrying out the stacking procedure after smoothing to coarser spatial resolutions. We find that increasing the size of the beam beyond 60~kpc does not result in a statistically-significant increase in the average \hii\ emission signal. We thus find no evidence that the average \hii\ emission from star-forming galaxies at $\langle z \rangle \approx 1.3$ arises from a region larger than 60~kpc. 
	
	The narrow beam of our GMRT observations implies that the effect of ``source confusion'' on our results is negligible. For example, \citet{Chowdhury20} estimated that the \hii\ emission from companion galaxies is very unlikely to lie within both the 60~kpc beam and within $\approx \pm 200$~\kmps\ of the target galaxies. They found that companion galaxies  contribute $\lesssim 2\%$ to  the stacked \hii\ emission signal.
	
	\begin{figure*}
		\centering
		\includegraphics[width=0.4\linewidth]{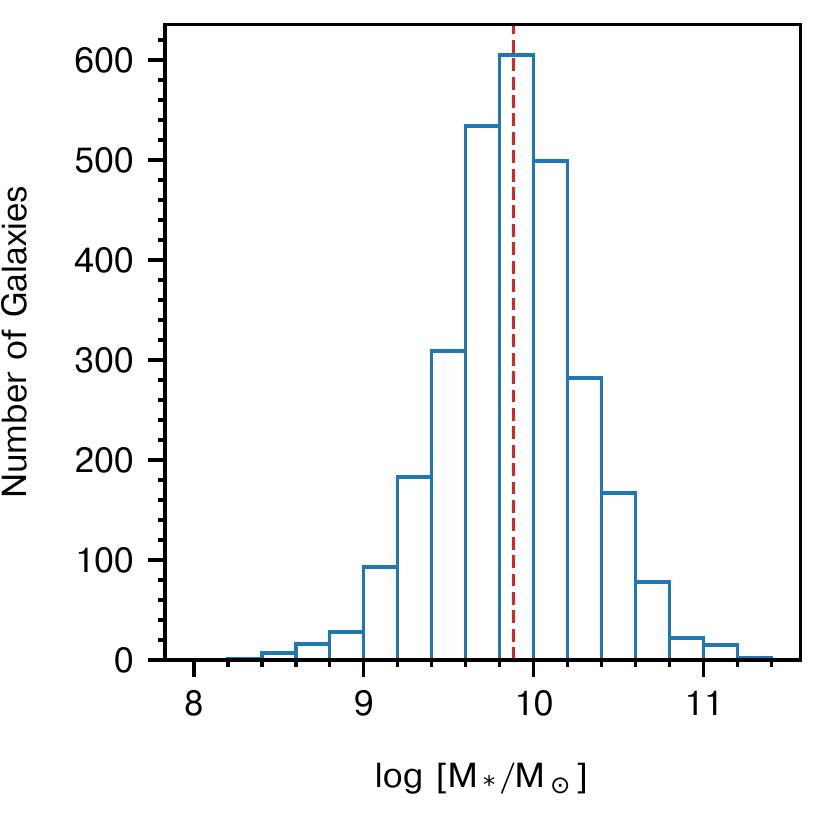}
		\caption{ The stellar mass distribution of our sample of  2,841 blue star-forming galaxies at $z\approx1.3$. The dashed red vertical line shows the median stellar mass of the sample.}
		\label{fig:stellarmass}
	\end{figure*}
	
     The 2,841 blue star-forming galaxies at $z\approx 1.3$ have stellar masses in the range $\Ms  \approx10^{8}-10^{11.5}\ \Msun$, with a mean stellar mass of $\langle \Ms \rangle=1.2\times10^{10}\ \Msun$.\footnote{All stellar masses and SFRs assume a Chabrier initial mass function.}\footnote{The individual stellar masses of the galaxies were inferred from a relation between the $\rm (U-B)$ color and the ratio of the stellar mass to the B-band luminosity, calibrated at $z\approx1$ using stellar masses estimated from K-band observations of a subset of the DEEP2 sample \citep{Weiner09}. Figure \ref{fig:stellarmass} shows the distribution of stellar masses of our sample galaxies. We split the 2,841 galaxies into two sub-samples, with stellar masses above and below their median stellar mass of $\Ms=7.6\times10^{9}\ \Msun$, and separately stack the \hii\ cubes of the galaxies in each sub-sample. The stacked \hii\ emission signal is indeed detected separately in the two stellar-mass sub-samples, albeit at lower statistical significance, at $\approx (3.4-3.7)\sigma$ level. We obtain average \hi\ masses of $\langle \MHI \rangle=(2.95\pm0.87) \times 10^{10}\ \Msun$ and $\langle \MHI \rangle=(3.23\pm0.88) \times 10^{10}\ \Msun$ for the lower-stellar mass ($\langle \Ms \rangle = 4 \times 10^9\ \Msun$) and higher-stellar-mass ($\langle \Ms \rangle = 2 \times 10^{10} \ \Msun$) sub-samples, respectively.  }

	Finally, we note that the DEEP2 survey is complete for galaxies with rest-frame B-band magnitude $\MB\le-21$ \citep{Willmer06,Newman13}. We hence stacked the \hii\ spectra of the 2,254 galaxies of our sample with $\MB\le-21$, following the above procedure. This too yielded a clear detection (with $\approx 4.4\sigma$ significance) of the average \hii\ signal, but from a complete sample of galaxies. Integrating over the detected \hii\ line profile, we find that blue star-forming galaxies with $\MB\le-21$ at $z\approx1.3$ have an average \hi\ mass of $\langle \MHI \rangle=(3.04\pm0.69)\times10^{10}\ \Msun$; this average \hi\ mass is consistent with that obtained for the full sample of galaxies.
		
	\section{Stacking the rest-frame 1.4~GHz Continuum Emission: The Radio SFR}
	\label{sec:sfr}
	\begin{figure}
		\centering
		\includegraphics[width=\linewidth]{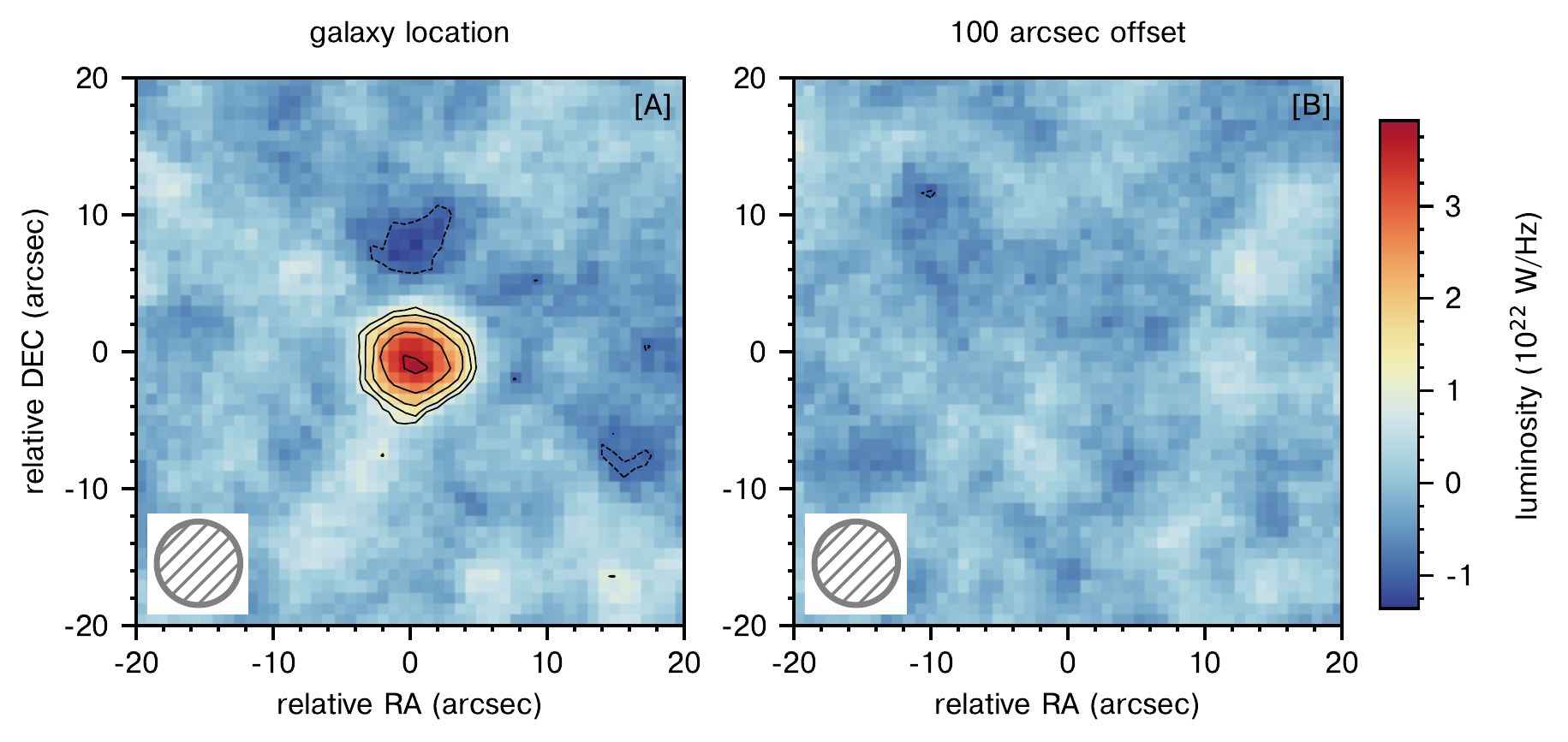}
		\caption{[A] The average rest-frame 1.4~GHz luminosity of our 2,841 star-forming galaxies, obtained from a weighted-median stack of their rest-frame 1.4~GHz emission. A clear ($\approx 12\sigma$ significance) detection of the median 1.4~GHz emission of the galaxies can be seen in the image. [B] The result of the offset stack, a weighted-median stack at locations $100''$ offset from the above galaxies: no signal is seen in this image. The circle at the bottom left of each panel shows the $6.1"$ resolution of the images. The contours are at $(-3.0, 3.0, 4.2, 6.0, 8.4, 12.0) \times \sigma$, with dashed negative contours.}
		\label{fig:contstack}
	\end{figure}
	
	The rest-frame 1.4~GHz luminosity (${\rm L_{1.4 GHz}}$) of a galaxy can be used to determine its total SFR, via the far-infrared-radio correlation  \citep[e.g.][]{Yun01,Pannella15,Bera18}. We initially convolved all our 610~MHz continuum images of the DEEP2 sub-fields (see Fig.~\ref{fig:cont}) to a common beam of size $6.1'' \times 6.1''$. For sub-fields 31/32 \& 32/33 of \citet{Kanekar16}, we used the images of \citet{Bera18}, also convolved to the same beam; however, for sub-fields 21 $\&$ 22, we use the new, deeper images of Fig.~\ref{fig:cont}. From the convolved images, we extract cutouts of size $25'' \times25''$ around each of our 2,841 galaxies. We also extract identically-sized cutouts $100''$ away from each galaxy, to test for the presence of systematic effects.  For each galaxy, we convert the flux density $S_\nu$ at each pixel of its sub-image to the corresponding rest-frame 1.4~GHz luminosity, assuming a typical spectral index of $\alpha=-0.8$, with $S_{\nu} \propto \nu^{\alpha}$ \citep{Condon92}. For the galaxies of our sample, the  observing frequencies are very close to rest-frame 1.4~GHz; our results are hence insensitive to the exact choice of spectral index.  
	
	We used a median-stacking approach \citep[e.g.][]{White07} to estimate the average ${\rm L_{1.4 GHz}}$ of the sample. This was done by computing the weighted-median of each pixel across the sample, with the weights being the same as used while stacking their \hii\ emission. Fig.~\ref{fig:contstack}[A] shows the average ${\rm L_{1.4 GHz}}$ emission from the 2,841 galaxies; the image shows a clear detection of the stacked continuum emission, at $\approx12\sigma$ statistical significance. No evidence for systematic effects is discernible in the offset-stack image of Fig.~\ref{fig:contstack}[B]. We obtain a median rest-frame 1.4~GHz luminosity of ${\rm L_{1.4 GHz}}=(3.93\pm0.30) \times 10^{22}$~W~Hz$^{-1}$. Using the SFR calibration of \citet{Yun01},  this implies a median SFR of $(14.5 \pm 1.1) \ \Msun$~yr$^{-1}$ for the 2,841 galaxies of our sample.

	\section{Discussion}
	
	Our detection of the stacked \hii\ emission signal of Fig.~\ref{fig:signal} is only the second, after \citet{Chowdhury20}, of \hii\ emission at $z\gtrsim 1$! The observations reported here used receivers, signal path, and correlator that were different from the system used by \citet{Chowdhury20}, and, of course, were affected differently by any time-variable RFI. The present detection of the \hii\ emission signal, and the measurement of the average \hi\ mass of star-forming galaxies at $z\approx 1.3$, are hence important independent confirmations of the results
	of \citet{Chowdhury20}.
	
	Our sample of 2,841 blue star-forming galaxies at $z\approx 1.3$  has a mean stellar mass of $\langle \Ms \rangle=1.2\times10^{10}\ \Msun$. Our radio-derived average SFR estimate of $(14.5 \pm 1.1) \; \Msun$~yr$^{-1}$ then implies that the average properties of our galaxies are consistent with those of the star-forming main-sequence at $z \approx 1.3$ \citep[e.g.][]{Whitaker12,Leslie20}. 
	
	Our measurement of a mean \hi\ mass of $\langle \MHI \rangle=(3.09\pm0.61)\times10^{10}\ \Msun$ implies that the typical \hi\ masses of main-sequence galaxies at $z\approx 1.3$ are significantly larger than their stellar masses, with an average \hi-to-stellar mass ratio of $2.6\pm0.5$. This is very different from the situation in the local Universe, where the \hi-to-stellar mass ratio is only $\approx 0.4$, for blue star-forming galaxies with a similar stellar mass distribution in the xGASS sample \citep[e.g.][]{Saintonge17,Chowdhury20}.\footnote{We estimate the average \hi\ mass to average stellar mass ratio of the xGASS sample by taking a ratio of the weighted-mean \hi\ mass and the weighted-mean stellar mass of star-forming galaxies, with NUV-$r<4$, in the xGASS sample; the weights for each xGASS galaxy was chosen such that their effective stellar mass distribution is identical to that of our 2,841 galaxies at $z\approx1.3$. We note that xGASS contains galaxies with $\Ms >10^9\ \Msun$ while our sample goes down to $\Ms \approx 10^8\ \Msun$ but this has a negligible effect on the comparison since only $\approx2\%$ of our sample of 2,841 galaxies at $z\approx1.3$ have $\Ms < 10^9\ \Msun$.} 
	
	 We measure an average \hi\ mass of $\langle \MHI \rangle=(2.95\pm0.87) \times 10^{10}\ \Msun$ and $\langle \MHI \rangle=(3.23\pm0.88) \times 10^{10}\ \Msun$ in the two stellar mass sub-samples, with average stellar masses of $4\times10^9\ \Msun$ and $2\times10^{10}\ \Msun$, respectively. This implies average \hi\ mass to stellar mass ratios of $7.4\pm2.2$ and $1.61 \pm 0.44$, in the lower- and higher-stellar mass samples, respectively, again for blue star-forming galaxies at $z\approx1.3$. While this tentatively suggests (at $\approx 2.6\sigma$ significance) that galaxies with low stellar masses are more gas-rich than those with high stellar masses \citep[as has been seen in the local Universe; e.g.][]{Catinella18}, our current signal-to-noise ratio is not sufficient to determine the dependence of the \hi-to-stellar mass ratio on the stellar mass.

	CO emission studies of similar main-sequence galaxies at $z \approx 1-2$ have shown that their molecular gas mass is comparable to the stellar mass \citep[e.g][]{Tacconi20}.  Our results thus indicate that the atomic gas content of star-forming galaxies at this epoch is larger than their molecular gas content. Further, at this epoch, star-forming galaxies contain nearly four times more cold gas, including both atomic and molecular gas, than stars, very different from the local Universe.
	
	The \hi\ depletion timescale, t$_\textrm{dep,H{\textsc{i}}} \equiv \textrm{M}_\textrm{H{\textsc{i}}}/\textrm{SFR}$, provides an estimate of how long a galaxy's \hi\ can fuel its star-formation (with an intermediate conversion to $\htwo$) at its current SFR. Combining our estimates of the average SFR and the average \hi\ mass yields an average \hi\ depletion timescale of $\langle{\rm t_{dep,H{\textsc{i}}}}\rangle = 2.13 \pm 0.45$~Gyr for star-forming
	galaxies at $z \approx 1.3$. This is similar to the \hi\ depletion timescale of $1-2$~Gyr obtained by \citet{Chowdhury20}, for star-forming galaxies at $z\approx 1.0$. In the local Universe, the \hi\ depletion time in main-sequence galaxies with a stellar mass distribution similar to that of our sample is
	$\approx 7$~Gyr \citep[e.g.][]{Saintonge17}. Our results \citep[and those of][]{Chowdhury20} thus indicate that the \hi\ depletion time changes significantly from $z \approx 1.3$ to $z = 0$, by a factor of $\gtrsim 3.5$.
	
	In local galaxies, the $\htwo$ depletion timescale, $\rm t_{dep,H_2}$, is relatively short, $\lesssim 1$~Gyr \citep[e.g.][]{Tacconi20}, far smaller than the \hi\ depletion timescale \citep[$\approx 5-10$~Gyr, depending on the stellar mass; e.g. ][]{Saintonge17}. Hence, star-formation activity in local galaxies is not typically limited by the \hi\ depletion timescale, as there is sufficient time for the \hi\ to be replenished, via either accretion from the circumgalactic medium (CGM) or mergers. As such, main-sequence galaxies in the local Universe can  continue to quiescently form stars at their current SFR for $\approx 5-10$~Gyr without the need for fresh gas accretion, as long as there is efficient conversion of \hi\ to $\htwo$ (on timescales shorter than $\rm t_{dep, H_2}$). Conversely, main-sequence galaxies at $z \approx 1.3$ can sustain their current SFR for only $\approx 2$~Gyr, unless their \hi\ reservoir is replenished. The short \hi\ depletion timescale could thus be a bottleneck for continued star-formation activity, in the absence of the acquisition of new \hi\ from CGM accretion or mergers. We note that the \hi\ depletion time of $\approx 2$~Gyr is comparable to the timescale on which the cosmic SFR density is observed to begin its steep decline. Our results are thus consistent with the hypothesis of \citet{Bera18} and \citet{Chowdhury20} that the quenching of star-formation activity at $z < 1$ may arise due to insufficient gas accretion, resulting in a paucity of neutral gas to fuel further star-formation.

	In passing, we note that \citet{Chowdhury20} obtained an average \hi\ mass of $\langle \MHI\rangle=(1.19\pm0.26)\times 10^{10}\  \Msun$
	for 7,653 star-forming galaxies at $z=0.74-1.45$, also in the DEEP2 fields. While this average \hi\ mass appears to be lower than our estimate, the wider redshift coverage of \citet{Chowdhury20} implies that the stellar mass and luminosity distributions of the two samples are not the same. For example, the average absolute B-band magnitude of the 7,653 galaxies of \citet{Chowdhury20} is $\langle \MB \rangle = -20.9$, while that of the 2,841 galaxies in our sample is $\langle \MB \rangle = -21.4$; i.e. the former sample contains more faint galaxies than our sample. We hence restricted the comparison to galaxies with $\MB\le-21$, the completeness limit of the DEEP2 survey \citep{Newman13}. The 3,499 such galaxies  of \citet{Chowdhury20} yield $\langle \MHI\rangle = (1.70\pm0.43)\times10^{10} \ \Msun$ at $\langle z \rangle \approx 1.0$, while we obtain $\langle \MHI \rangle=(3.04\pm0.69)\times10^{10}\ \Msun$ at $\langle z \rangle \approx 1.3$, consistent at $\approx 1.6\sigma$ significance. We thus find no evidence for redshift evolution in the average \hi\ mass from $z \approx 1.3$ to $z \approx 1.0$.

	\section{Summary}
	We report a $\approx 400$-hour GMRT \hii\ emission survey of the DEEP2 fields, carried out using the original GMRT 610~MHz receivers and the legacy narrowband GMRT correlator, which (including 60~hours of earlier data with the same system), has yielded only the second detection of redshifted \hii\ emission from galaxies at $z \gtrsim 1$. We stacked the \hii\ emission from 2,841 blue star-forming, main-sequence galaxies at $z = 1.18-1.39$ to obtain a $5.0\sigma$-significance detection of the average \hii\ emission signal of the galaxies. This implies an average \hi\ mass of $\langle \MHI\rangle=(3.09\pm0.61)\times10^{10}\  \Msun$ for blue star-forming galaxies at $\langle z\rangle = 1.3$. We obtain an  \hi-to-stellar mass ratio of $2.6\pm0.5$, significantly higher than that of similar galaxies in the local Universe. We use our radio-continuum images to infer an average SFR of $(14.5 \pm 1.1) \ \Msun/\textrm{yr}$ for the same 2,841 galaxies and combine this with our average \hi\ mass estimate to obtain an average \hi\ depletion timescale of $(2.13 \pm 0.45)$~Gyr, consistent with the results of \citet{Chowdhury20}, and far shorter than the \hi\ depletion timescale of galaxies with similar stellar masses in the local Universe.  The short \hi\ depletion timescale is comparable to the timescale on which the cosmic SFR density declines after its peak at $z\approx1-3$. Our results are thus consistent with the hypothesis that the availability of \hi\ in galaxies at $z \approx 1$ may been a critical bottleneck in quenching their star-formation, which led to the observed decline in the cosmic SFR density at $z < 1$. 
	
	\acknowledgments
	We thank the staff of the GMRT who have made these observations possible. The GMRT is run by the National Centre for Radio Astrophysics of the Tata Institute of Fundamental Research. NK acknowledges support from the Department of Science and Technology via a Swarnajayanti Fellowship (DST/SJF/PSA-01/2012-13). AC, NK, $\&$ BD also acknowledge the Department of Atomic Energy for funding support, under project 12-R\&D-TFR-5.02-0700. 
	
    \software{CASA \citep{McMullin07},   
      calR \citep{calR},  AOFLAGGER \citep{Offringa12}, 
      astropy \citep{astropy:2013,astropy:2018}}
\bibliographystyle{aasjournal}

\end{document}